\documentclass{aa} 

\usepackage{natbib}
\bibpunct{(}{)}{;}{a}{}{,} 

\usepackage{graphicx}

\usepackage{amssymb,amsmath}
\usepackage{delarray}
\usepackage{mathrsfs}
\usepackage{xcolor}
\usepackage[varg]{txfonts}

\begin{document} 


\title{Observations on spatial variations of the Sr~{\sc i} 4607~\AA~ scattering polarization signals at different limb distances
with ZIMPOL}

\author{Sajal Kumar Dhara\inst{1}
         \and Emilia Capozzi\inst{1}
         \and Daniel Gisler\inst{1,}\inst{2}
         \and Michele Bianda\inst{1}
         \and Renzo Ramelli\inst{1}
         \and Svetlana Berdyugina\inst{2}
         \and Ernest Alsina\inst{1}
         \and Luca Belluzzi\inst{1}
       }
\institute{Istituto Ricerche Solari Locarno (IRSOL), 6605 Locarno-Monti, Switzerland,\\  \email{sajal@irsol.ch}   
           \and Kiepenheuer-Institut f\"ur Sonnenphysik (KIS),
            Sch\"oneckstrasse 6, 79104 Freiburg, Germany\\
          }

\titlerunning{Spatial variations of the Sr~{\sc i} 4607~\AA~ scattering polarization signals}
\authorrunning{Dhara et al.}

\abstract
{
The Sr~{\sc i} 4607~\AA\ spectral line shows one of the strongest scattering 
polarization signals in the visible solar spectrum. The amplitude of this polarization signal is expected to 
vary at granular spatial scales, due to the combined action of the Hanle effect and  the local  
anisotropy of the radiation field. Observing these variations would be of great interest because it would 
provide precious information on the small-scale activity of the solar 
photosphere. At  present, few detections of such spatial variations have been reported. This is due to the difficulty of these 
measurements, which require  combining high spatial  ($\sim$ 0.1"), spectral  ($\leq$ 20 m\AA), and
temporal resolution  (< 1 min) with increased polarimetric sensitivity  ($\sim$ 10$^-$$^4$).
}
{
We aim to detect spatial variations at granular scales of the 
scattering polarization peak of the Sr~{\sc i} 4607~\AA\ line at
different limb distances, and to study the correlation with the continuum 
intensity.
}
{
Using the Zurich IMaging POLarimeter (ZIMPOL) system mounted at the GREGOR 
 telescope and spectrograph in Tenerife, Spain, we carried out spectro-polarimetric 
measurements to obtain the four Stokes parameters in
 the Sr~{\sc i} line at different limb 
distances, from $\mu=0.2$ to $\mu=0.8$, on the solar disk.
}
{
Spatial variations of the scattering polarization signal in the 
Sr~{\sc i} 4607~\AA\ line, with a spatial resolution of about 0.66", are clearly observed at every $\mu$. 
The spatial scale of these variations is comparable  to the granular size. 
A statistical analysis reveals that the linear scattering polarization amplitude in this Sr~{\sc i} spectral line 
 is positively correlated with the intensity in the continuum, corresponding to the granules,
at every $\mu$.
}
{}


\maketitle

\section{Introduction}\label{sec:sec1}

The Sr~{\sc i} 4607~\AA~ photospheric line shows one of the strongest scattering 
polarization signals \citep[above 1\% at $\mu$ =  Cos($\theta$) = 0.1, where $\theta$ is 
the heliocentric angle; see e.g.,][]{Stenflo1997,Gandorfer2002}  
in the visible solar spectrum. 
This signal is sensitive to the Hanle effect  
and can be used as an excellent diagnostic tool for investigating
weak turbulent magnetic fields in quiet photospheric regions \citep{Faurobert1993,Faurobert2001}.
By solving the radiative transfer problem of scattering 
line polarization in a realistic three-dimensional (3D) hydrodynamical model of the solar
photosphere, \cite{Bueno2007} foresee appreciable small-scale
spatial variations of the amplitude of the scattering polarization signals
in the Sr~{\sc i} 4607~\AA~line,
 provided that granulation is resolved.
The origin of such variations is related to variations in the magnetic 
field present in the granules or the intergranular lanes, and to the
local variations in the anisotropy of the radiation field.
Unfortunately, detecting these spatial variations is challenging. 
Observations with the polarimetric sensitivity required to detect this scattering polarization signal, generally suffer
from insufficient spatial and temporal resolution \citep{Bueno2007}.
A further difficulty is due to the fact that scattering polarization signals decrease when moving from the solar limb to the disk
center, while the granulation can be clearly resolved only if the
observed region is sufficiently far from the limb.

\cite{Malherbe2007} performed spectro-polarimetric observations of the Sr~{\sc i} 4607~\AA~ line at 40" ($\mu$=$\sim$0.3)
inside the solar limb
with a spatial resolution of about 0.6''. Their observations
reported that intergranules are generally less polarized than
the granules, thus suggesting stronger magnetic fields
in the intergranular lanes.  Their findings perhaps hint at
a Hanle effect acting in the intergranules (meaning in higher
magnetic field regions), in agreement with simulations carried out by \cite{Bueno2004}.
 \cite{Bianda2018} carried out spectro-polarimetric observations at $\mu$=0.3 close to the east limb
using the Zurich IMaging POLarimeter \citep[ZIMPOL:][]{Ramelli2010} at the
GREGOR telescope  \citep{Schmidt2012,Ramelli2014}. Their observations also
revealed small-scale spatial variation of the scattering polarization signal of
the Sr~{\sc i} 4607~\AA~line, and confirmed that the amplitudes are generally higher at the center of the 
granules than in the intergranular lanes , meaning the polarization signal is positively correlated to the continuum intensity.

Recently, \cite{Aleman2018} performed 3D radiative transfer calculations in a high-resolution
magneto-convection model, finding an anticorrelation between
the amplitude of the theoretical scattering polarization signal of the Sr~{\sc i} 4607~\AA~line and
the continuum intensity at all on-disk positions.
Their work also shows how the predicted amplitudes  and 
spatial variations are modified 
after degrading the  signal-to-noise ratio (S/N) and the spectral and spatial
resolutions of the simulated observations.
Stokes filtergraph observations of the Sr~{\sc i} 
4607~\AA~line have been carried out by \cite{Zeuner2018} with the Fast 
Solar Polarimeter  \citep[FSP:][]{Feller2014,Iglesias2016} mounted on Telecentric Etalon SOlar Spectrometer (TESOS)
at the German Vacuum Tower Telescope  \citep[VTT:][]{Soltau1981}, Tenerife,
Spain.  The TESOS Fabry-Perot tunable filtergraph has a full-width at half maximum (FWHM) of about 25 m\AA.
These observations performed  at $\mu$ = 0.6  toward the north
solar limb  for field of view (FOV) 20"$\times$20", sampled
 with 0.16'' pixel$^-$$^1$. They scanned the Sr~{\sc i} absorption line at 
 five wavelength positions with a step of 30~m\AA~ and 2.5 s integration time at each position.
Their results shows an anti-correlation between
the amplitudes of the Stokes Q/I signals and the continuum intensity. The observed larger scattering polarization signals 
in the intergranules is in qualitative agreement with theoretically predicted by \cite{Bueno2007} and \cite{Aleman2018}.

As a further step along this line of research, we carried out an observing campaign during June 2018 at the GREGOR solar telescope 
 in Tenerife, using the ZIMPOL polarimeter system. Our main aim was to measure small-scale spatial variations
of the Sr~{\sc i} 4607~\AA~scattering polarization  signal, at different limb distances on the solar disk.
 We describe the observations with the GREGOR telescope and the data
 reduction process  in Sect. 2. We study the possible presence of correlations between the amplitudes of
 the scattering polarization signals
 and the continuum intensity,  Sect. 3, which we have used as an
 indicator of the locations of granules and intergranular lanes, at the various limb
distances considered.
 A discussion of our results is presented in  Sect. 4.

\section{Observations with GREGOR}\label{sec:sec2}

Our observational campaign at the GREGOR telescope was held between 13 June 2018 and 27 June 2018. 
The ZIMPOL system was installed at the GREGOR spectrograph.   
The polarimeter analyzer consists of a double ferroelectric crystal modulator, 
followed by a linear polarizer. 
It was mounted in front of the spectrograph slit. 
The ferroelectric crystal  modulates the signals at the frequency of 1 kHz, thus allowing minimal 
spurious effects induced by intensity variations due to the seeing. 
A synchronous demodulation can be obtained with the ZIMPOL camera,
which has a masked charge-coupled device sensor equipped with cylindrical microlenses \cite[for details see][]{Ramelli2010}.
During our observation we also used the image derotator, which  allows to maintain the slit orientation with respect to the rotating
solar image. This is thus an
advantage for performing long exposure measurements.
 However, the disadvantage of using the image derotator is that it introduces
  extra variable instrumental polarization. Thus, more calibrations are needed during the observations. At noon, we did not perform observations
as the Sun is near to zenith and the image rotation is  rapid (>80" s$^-$$^1$). 
By contrast, it is quite stable  (< 20" s$^-$$^1$) during the early morning and the evening.

\begin{table*}
\centering
\footnotesize

\caption{Details of observations performed during campaign. 
The spectrograph slit was placed parallel to the nearest limb of the Sun during measurements.
}
\begin{tabular}{|l|l|l|l|l|l|}
 \hline
&&&&&\\        
  Limb             &  Date of      & Slit  position on   & Total duration         &  no. of frames        & Fried parameter ($r_{0}$)    \\
  distances ($\mu$)&  observations & the solar disk      &  (min)                 & acquired              & during measurements       \\

\hline  
&&&&&\\
 
  0.2   &  27  June      &  near W limb                    &  8.9    & 80   &  5 cm, a small plage region\\
  &&&&                                                                      & is used to lock the AO. \\
    0.38  &   23  June   &   near E limb, close to sunspot &  3.4    & 30   &  4 cm.\\

  0.4   & 16  June       &  near N limb                    &  16.5   & 148  &  5 -- 10 cm. \\
  0.44  & 23  June       &  near E limb, close to sunspot  &  6.7    & 60   &  4 -- 7 cm.\\

  0.5   &  16  June      &   near N limb                   &  16.5   & 148  &  4 -- 7 cm,\\
  &&&&                                                                      &  sometimes AO was unstable. \\
  0.6   &  27  June      &   near W limb                   &  8.9    & 80   &  4 -- 5 cm.\\
    &&&&                                                                     &  sometimes AO was unstable. \\

  0.7   &   21  June     &   near N limb                   &  16.5   & 148   &  3 -- 5 cm, \\
 &&&&                                                                        & sometimes AO was unstable.\\
  0.8   &   16  June     &  near N limb                    &  14    & 125      &  5 -- 10 cm. \\
        
 disk center &  21  June                                   &  --  &  16.5   & 148      &  5 -- 7 cm. \\

\hline
\end{tabular}
\label{Table:obs1}
\end{table*}

The observations were performed at  quiet regions of the Sun, 
at different limb distances on the solar disk
from $\mu$=0.2 to 0.8. The date, limb distances, and other details of
the observations are given in Table~\ref{Table:obs1}.
We used the adaptive optics (AO) system \citep{Berkefeld2016} for all measurements in order to
get stable observations for the required field-of-view (FOV) positions. 
Close  to  the disk  center  ($\mu$ = 0.8  to  0.4),  the granular  structures  were  used  by  
the   Shack-Hartmann wavefront sensor  to lock the AO system;  
while closer to the limb  ($\mu$ = 0.2), where the intensity contrast is low,  bright  plage  regions
present in the FOV were used  instead  during observations. 
The spectrograph slit was almost always placed  away from active regions, parallel to the nearest limb. The seeing quality 
fluctuated during our observations. 
Observations taken into account in the analysis are only
those with a spatial resolution better than 1'' for at least $\sim$ 2 min, during which granular and intergranular 
regions are well distinguished.
The observed Fried parameters during our measurements are listed in Table~\ref{Table:obs1}. 
The spectrograph slit covers a solar area of 0.3" (width of the slit) times 47" (length of the slit).
The obtained ZIMPOL image  has a spatial scale of $\sim$0.33" per pixel and it contains 140 pixels in the spatial direction. 
The estimated spatial resolution of the Stokes images, under
adequate seeing conditions, is found
to be  0.66", but it changes for different sets of measurements depending upon the seeing conditions.
The spectral resolution of our observation is of 10 m\AA.
We used the GREGOR polarimetric calibration unit \citep[][]{Hofmann2012}, which is mounted at the second focal point 
before any folding reflection, in order to avoid any significant instrumental polarization produced by the mirrors.

For all selected regions, we followed an observing procedure similar to the one adopted in the campaign of the previous year,
(see \cite{Bianda2018} for further details).
In brief, the observing procedure is the following. i) Acquisition of instrumental polarization measurements are used to calculate
special ZIMPOL camera 
timing parameters, which are needed to electronically compensate the large instrumental polarization 
 \citep[see][for more details]{Ramelli2014}.
ii) Polarimetric calibration is performed using the GREGOR polarimetric calibration unit including dark frame measurement. The acquired data are used to calculate the
demodulation matrix. 
iii) Flat fielding was taken by moving the telescope in quiet regions at the solar disk center. iv) Finally, scientific observations were performed in the selected region. The steps (ii) and (iii) are repeated again after the scientific observation.
We have taken series of frames at every observed region depending upon the seeing conditions. One single frame contains
the four modulated intensities to calculate the four Stokes parameters, but a series of four images are 
required to eliminate the spurious detector effects \citep{Gisler2005,Ramelli2014}. A typical exposure time for a single acquisition is
1 s. Therefore, with an additional overhead time, we can only reach a minimum time resolution around 7 s to digitize and transfer data.

\section{Data analysis and results}\label{sec:sec3}

 Data collected during the polarization calibration procedure before each measurement was 
 used to generate the calibration matrix, which was
 subsequently used to calculate the Stokes images acquired with ZIMPOL. Stokes intensity images were corrected for flat field, and taken 
 close to measurement time. 
 
 \begin{table*}
\centering
\footnotesize

\caption{For each measurement at a different $\mu$, we report the  approximate rms pixel noise of a single 
Stokes Q/I image (second column) as well as
of the image obtained after averaging various frames (third column) and the estimated spatial 
resolution (last column) for a single frame and averaged frame.
}
\begin{tabular}{|l|l|l|l|l|l|}
 \hline
    &&&&\\       
  Limb                 & Q/I rms noise (\%)  & Q/I rms noise (\%) (No. of frames    & \multicolumn{2}{c|}{Estimated spatial  resolution}     \\
 
  \cline{4-5}
  distances ($\mu$)    & (1 frame)  & averaged, total integration time)           &   single frame & averaged frame    \\ 

\hline  
 &&&& \\ 
  0.2          &  0.92 &  0.25 (12 frames,  1.34 min) &  0.66"- 0.99" &  0.66"- 0.99"\\ 
  0.38         &   0.72 &  0.16 (22 frames,  2.46 min) &  0.66"  &  0.66" - 0.99" \\ 

  0.4          &  0.73 &  0.19 (16 frames,  1.78 min) &  0.66"  & 0.99" \\
  
  0.44         &  0.69 &  0.18 (19 frames,  2.12 min) &  0.66"  & 0.66" - 0.99"\\

  0.5          &  0.76 &  0.13 (29 frames,  3.24 min) &  0.66"- 0.99"  &  0.66" - 0.99"\\ 
  
  0.6          &  1.00 &  0.22 (26 frames,  2.90 min)  &  0.66"- 0.99"  &  0.66" - 0.99"\\ 
  
  0.7          &  0.72 &  0.14 (33 frames,  3.68 min)  &  0.66" &  0.66" - 0.99" \\
  
  0.8          &  0.82 &  0.12 (62 frames,  6.98 min) &   0.66"  & 0.66" \\ 
        
 disk center   &  0.73 &  0.11 (50 frames,  5.58 min) &  0.66" &    0.66"\\

\hline
\end{tabular}
\label{Table:obs2}
\end{table*}

The statistical noise for a single Stokes image is comparable to the amplitude of the signals in the Sr~{\sc i} line.
The root mean square (rms) noises are calculated in a continuum area located in the Stokes Q/I image for each measurement. The rms pixel 
noise obtained from a single frame
are shown in Table~\ref{Table:obs2}. In order to improve the  S/N in the Stokes images, it
is necessary to average a set of frames. But, one also 
needs to continue  preserving the spatial and temporal resolutions. The fourth column of Table~\ref{Table:obs2} shows the number 
of frames over which
we averaged to improve the S/N for the measurements at different limb distances.

 \begin{figure*}
\centering
{\includegraphics[width=0.35\textwidth,clip=]{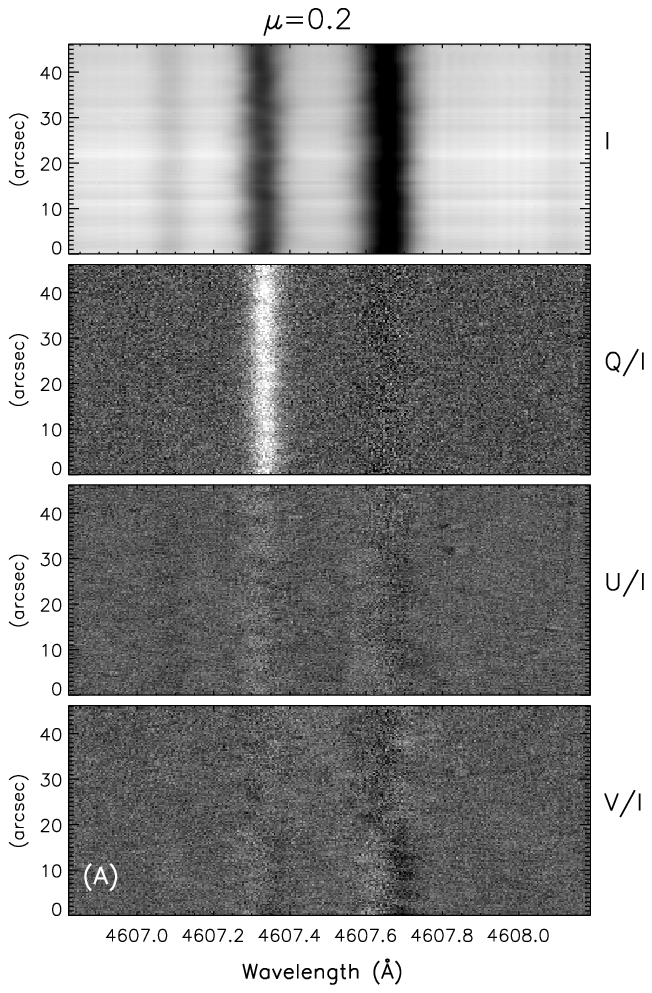}
   \hspace*{-0.04\textwidth}
 \includegraphics[width=0.35\textwidth,clip=]{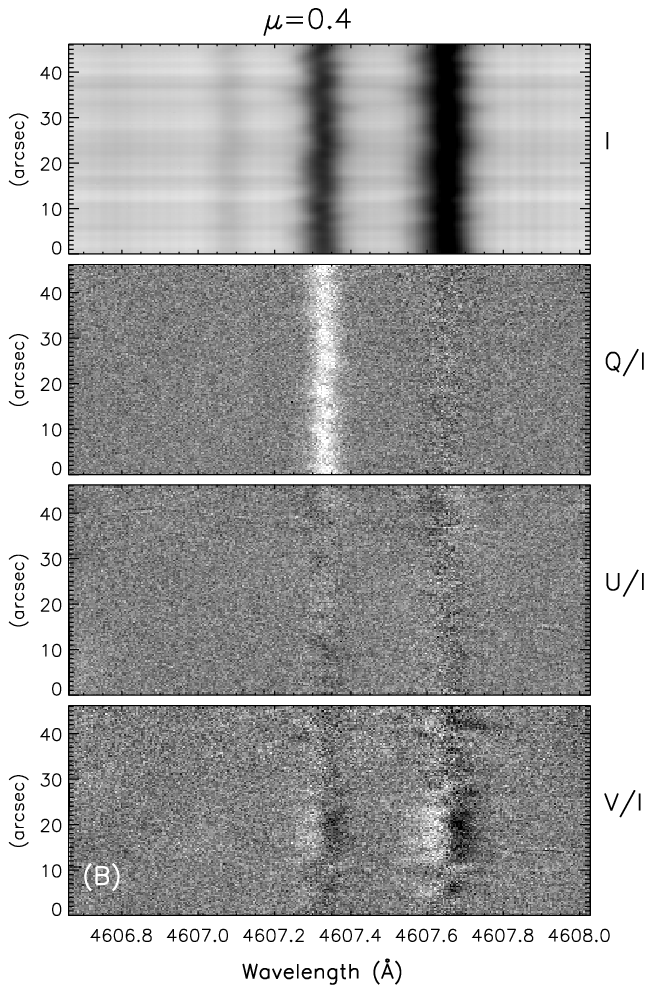}
  \hspace*{-0.04\textwidth}
 \includegraphics[width=0.35\textwidth,clip=]{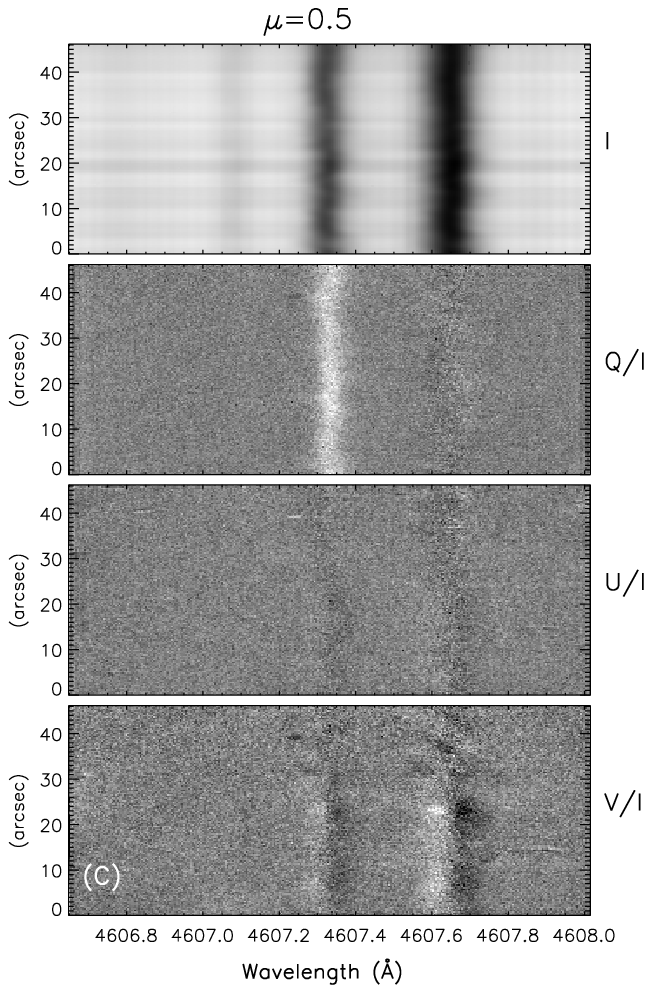}}
 {\vspace{-0.1\textwidth}
 \includegraphics[width=0.35\textwidth,clip=]{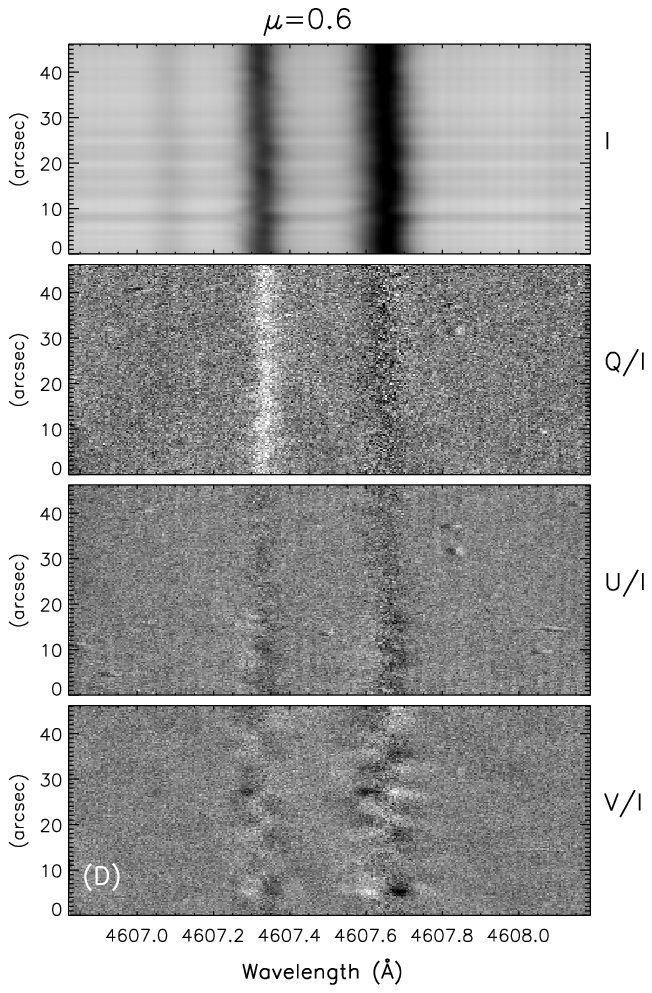}
   \hspace*{-0.04\textwidth}
 \includegraphics[width=0.35\textwidth,clip=]{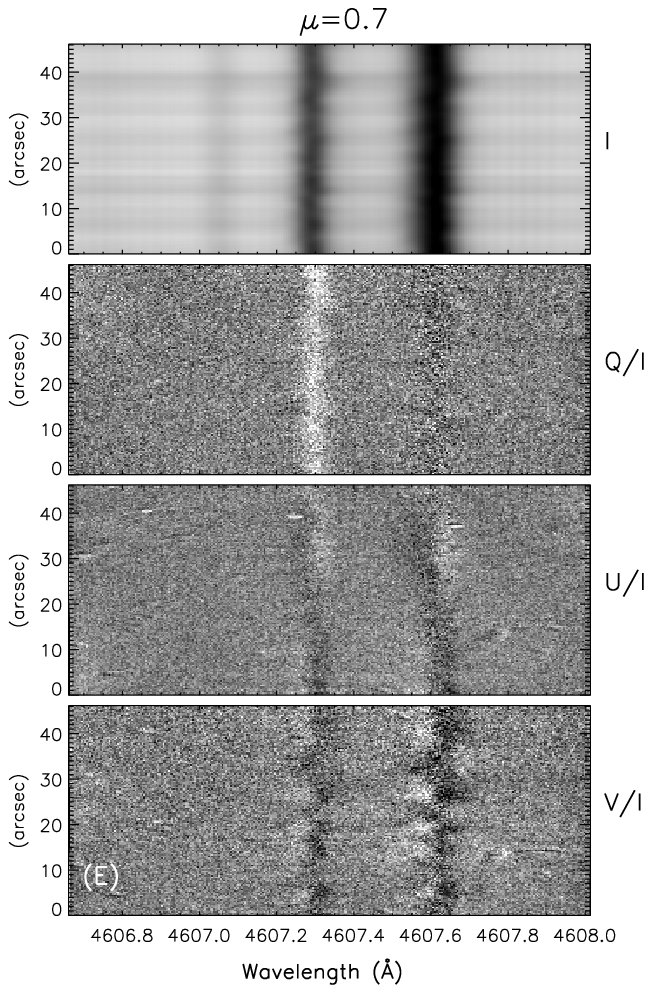}
  \hspace*{-0.04\textwidth}
 \includegraphics[width=0.35\textwidth,clip=]{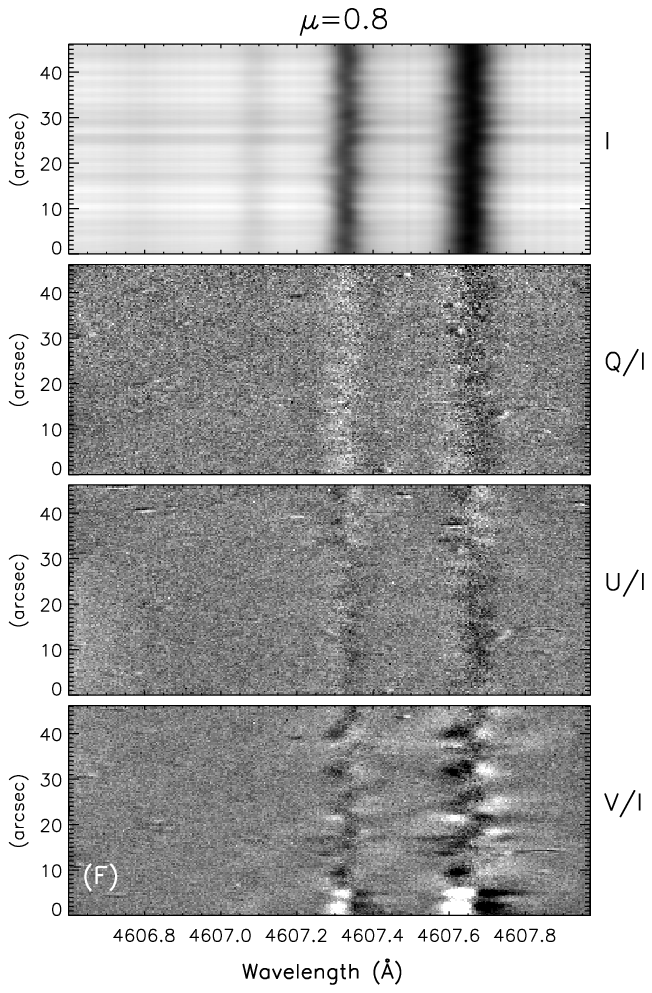}
 }
 \vspace{0.1\textwidth}
\caption{Stokes images of spectral interval around Sr~{\sc i} 4607~\AA~line. The spatial direction spans $\sim$47'' on the solar disk. 
The observed regions were at different limb distances ($\mu$ = 0.2, 0.4, 0.5, 0.6, 0.7, and 0.8). 
The slit was placed parallel to the nearest limb. 
The reference direction for positive Stokes Q is the tangent to the
nearest solar limb. These Stokes images are obtained after averaging over several frames (see Table~\ref{Table:obs2}).
The granulation pattern is visible in the intensity Stoke image,
especially in the continuum. The Q/I image shows the scattering polarization peak in the core of the Sr~{\sc i} line.
Spatial variations at granular
scales of this peak can be observed.
The typical
antisymmetric Zeeman patterns can be easily recognized in several Stokes V/I images.} 
\label{fig:stokes_image}
\end{figure*}

Figure~\ref{fig:stokes_image}
shows the Stokes images corresponding to different limb distances between $\mu$=0.2 and $\mu$=0.8, obtained after averaging 
over sequentially registered frames  (see the third column of Table~\ref{Table:obs2} for the corresponding number of frames averaged).
 In the Stokes I image one can
recognize intensity variations along the spatial direction due to the granulation. 
The Stokes Q/I image shows the scattering polarization
peak in the Sr~{\sc i} 4607~\AA~line. The polarization peak shows clear
spatial variations; their detection was indeed the main goal of our observations.
In the Stokes U/I image, there is no signal 
in the Sr~{\sc i} line core. 
In the Stokes V/I image, one can recognize
the typical patterns of the longitudinal Zeeman effect.

\subsection{ Solar evolution and image accumulation}\label{sec:sec4_1}

\begin{figure}
\centering
{\includegraphics[width=0.22\textwidth,clip=]{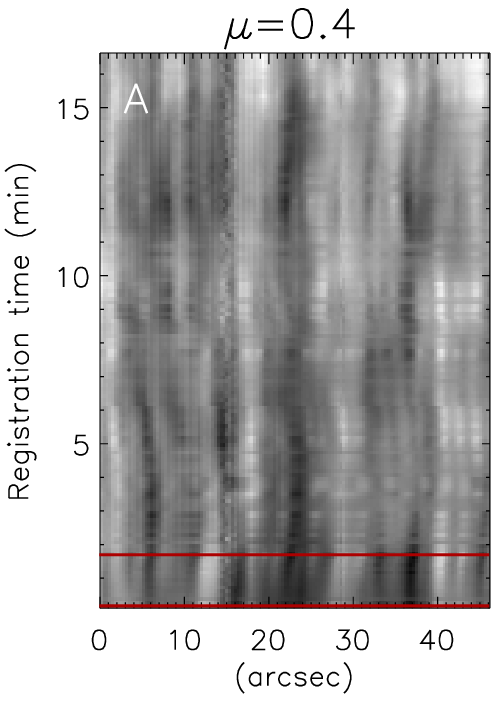}
 \hspace*{-0.01\textwidth}
\includegraphics[width=0.22\textwidth,clip=]{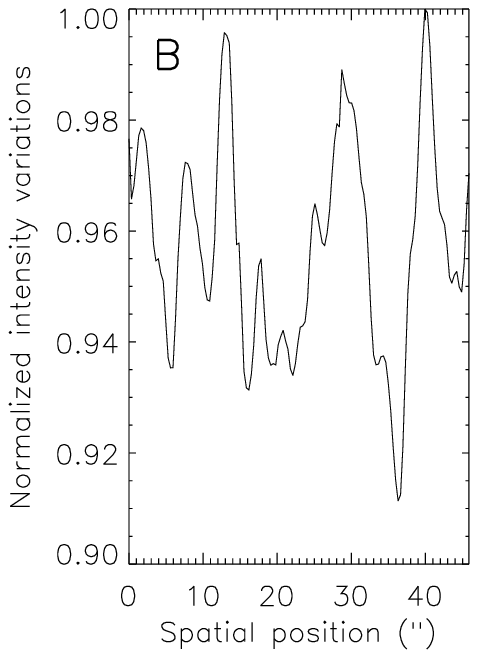}}
{\includegraphics[width=0.22\textwidth,clip=]{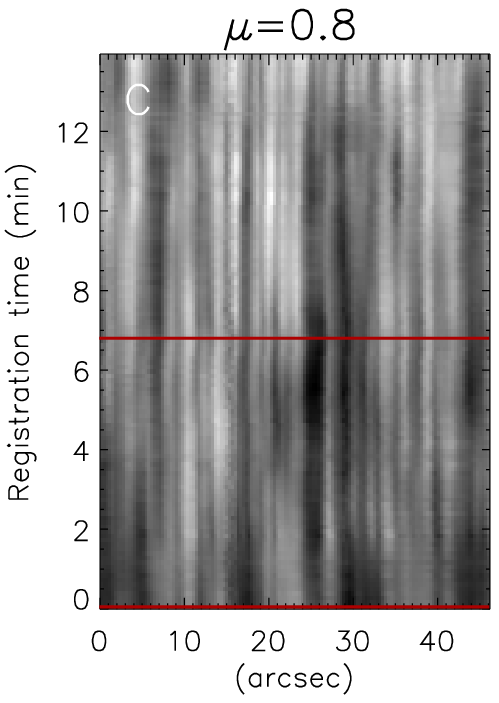}
 \hspace*{-0.01\textwidth}
\includegraphics[width=0.22\textwidth,clip=]{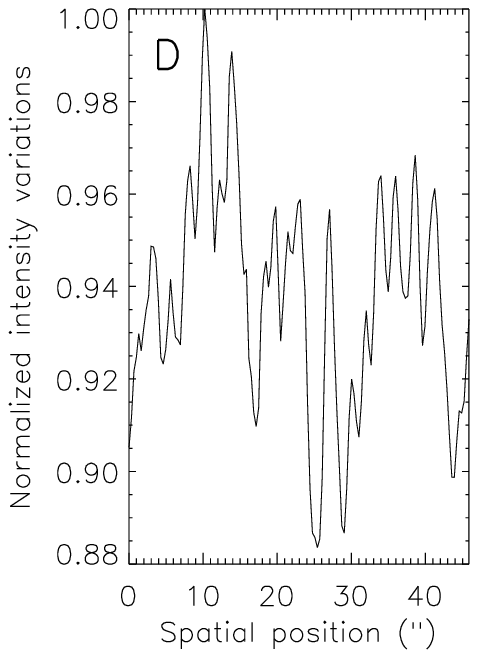}}
 \vspace{0.005\textwidth}
\caption{(A): Space-time map corresponding to observation at $\mu$ = 0.4. This map is generated by plotting  
the continuum intensity profile along the horizontal axis, obtained from 
each Stokes I images (of 148 total frames) along the spectrograph slit. Each profile was obtained at
6.7 seconds, so that the whole measurement shows a temporal evolution of the FOV of the spectrograph slit of 16.5 minutes.
(B): Normalized continuum intensity profile, obtained by averaged over 1.78 minutes from observation beginning.
The corresponding time interval
is indicated between the red horizontal lines in the space-time map. This profile is used as a parameter 
to indicate the granular and intergranular regions in Fig.~\ref{fig:scatter_plot}. 
(C): Space-time map corresponding to observation at $\mu$=0.8. (D): Normalized continuum intensity 
profile. This was obtained by averaging over 6.98 minutes from the
beginning of the observation (the corresponding time interval is indicated between red horizontal lines in
the space-time map).}
\label{fig:tme_space}
\end{figure}

In order to improve the S/N in the Stokes images, a
few subsequent frames were averaged over time. 
The number of frames considered in the average is selected by verifying through visual inspection that the
granular and intergranular regions are well distinguished over time and they have not significantly evolved during the corresponding
time interval. 
 We adopted this visual estimation of the solar granulation evolution to choose the appropriate time interval 
for performing the temporal average.
As an example, we consider the observation performed at $\mu$=0.4. The data series contains 148 frames corresponding to a
total observing time  of about 16.5 min. 
Taking into account the lifetime of granulation, we averaged 16 subsequent frames of the time series
(integration time of about 1.78 minutes), so as to reach a sufficient
 S/N. The ensuing Stokes images are shown in  Fig.~\ref{fig:stokes_image} (panel B).
To evaluate whether this 
temporal averaging introduced significant spatial degradation in our
data, we generated a space-time map of the frames seen by the spectrograph slit at a continuum wavelength.
 An example map is shown in Fig.~\ref{fig:tme_space} (panel A), where the continuum
intensity is presented in gray scale as a function of the spatial position along the slit (x-axis) and time (y-axis).
 This map shows the evolution of granulation over time. The observed granular structures remain unchanged within several minutes.
A single continuum intensity profile is shown as an example in Fig.~\ref{fig:tme_space} (panel B). This profile 
is obtained by averaging over 1.78 minutes 
(corresponding to 16 frames) from the beginning, 
and is normalized with respect to the maximum value. 
The variation of the intensity profile allows to distinguish the granular regions (larger intensity) and intergranular lanes (lower intensity).
 The final Stokes images are obtained 
  by averaging with a temporal window width of $\sim$1.78 minutes (16 frames), shifted in increments
of 16 frames (to avoid data overlap). A
few frames have been discarded due to poor seeing conditions during our observation. 
Those frames
are easily identified from the space-time map (Fig.~\ref{fig:tme_space}, panel A)  by visual inspection, when granular and 
intergranular regions are not well defined.

\begin{figure}[t]
\centering
{
\includegraphics[width=0.22\textwidth,clip=]{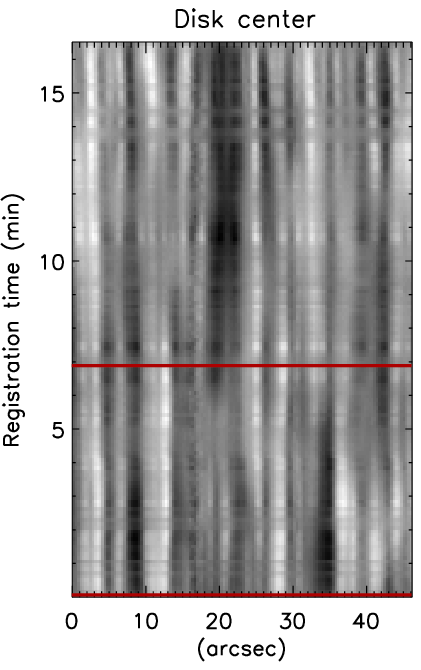}
 \hspace*{-0.02\textwidth}
\includegraphics[width=0.26\textwidth,clip=]{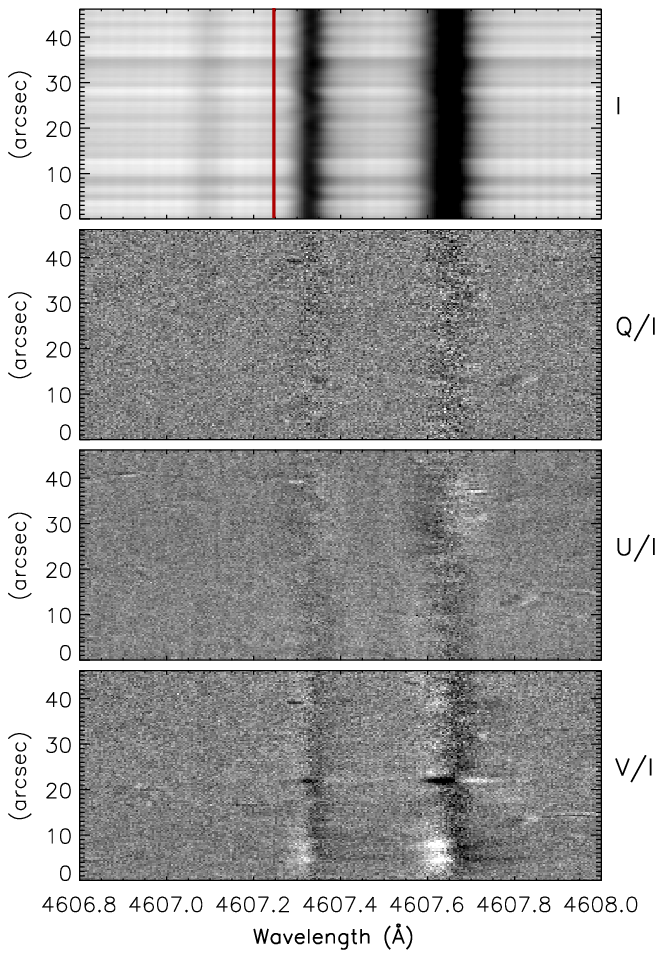}}
 \vspace{0.005\textwidth}
\caption{Left: Space-time map obtained at disk center. Right: Stokes images observed at  disk center.
These were observed in a spectral interval around the Sr~{\sc i} 4607~\AA~line, and averaged over $\sim$5.58 minutes (50 frames) from the beginning of the
observation (see the temporal interval between horizontal red lines
in the space-time map).  The red vertical line in the Stokes I image shows the wavelength position used to generate the space-time map.}
\label{fig:disk_center}
\end{figure}

Similar reduction processes have been applied to other sets of data obtained at different $\mu$.
The total number of frames acquired at every $\mu$ is given in Table~\ref{Table:obs1}.  
The number of averaged frames over time (meaning the temporal window width of each data set) is selected following  a visual estimation
of the solar granulation evolution discussed
above, allowing at the same time a good S/N
at each $\mu$.
These are mentioned in the third column of Table~\ref{Table:obs2}.
The temporal window width of each data set was chosen by visually inspecting the obtained space-time maps at every $\mu$. 
Close to the disk center, between $\mu$=0.5 and 0.8, the Q/I signal is low, as such, it is necessary to average over a longer time 
duration in order to sufficiently improve the S/N. While from $\mu$=0.4 to 0.2, the signal increases as a smaller 
number of averaged frames are required. The left panel of 
Fig.~\ref{fig:disk_center}  
shows the space-time map of images seen by 
the spectrograph slit at the disk center. One can clearly see the evolution of the granulation. 
Few frames have been discarded for final data reduction
due to poor seeing conditions. The right panel of Fig.~\ref{fig:disk_center} shows the obtained averaged Stokes images.
The Stokes Q/I and U/I images  present no signal in the Sr~{\sc i} 4607~\AA~line. 

\subsection{Stokes Q/I signal peak}\label{sec:sec4_2}

\begin{figure*}[t]
\centering
{
\includegraphics[width=0.8\textwidth,clip=]{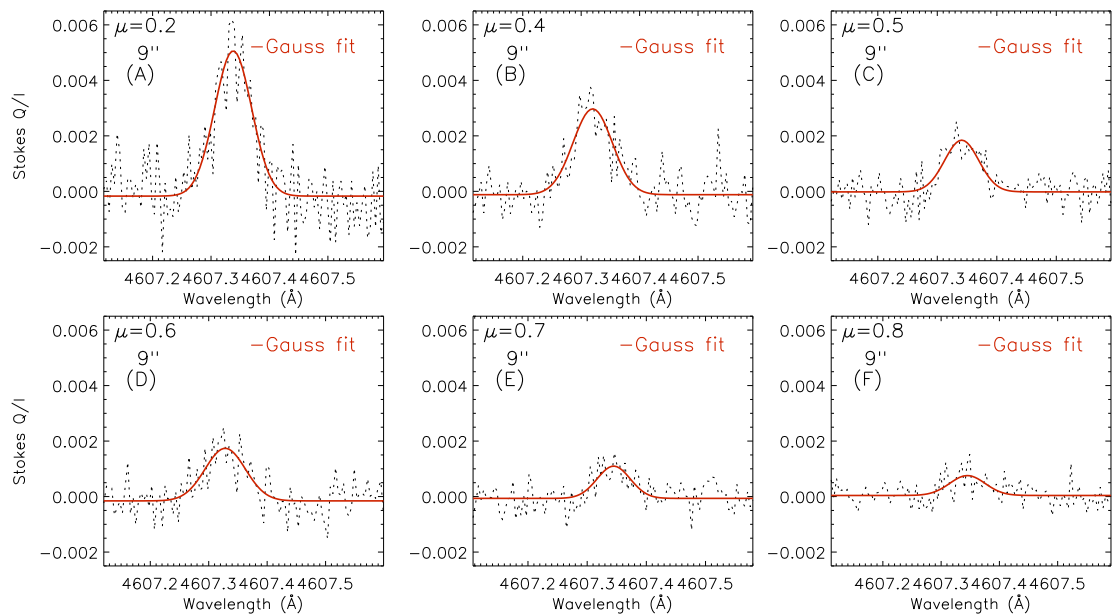}
}
\vspace{-0.05\textwidth}
\caption{Q/I profiles obtained at 9" spatial position for $\mu$ = 0.2 to 0.8. These profiles correspond to the Stokes Q/I images 
shown in Fig.~\ref{fig:stokes_image} (panels A--F). The red lines are the Gaussian fits to the profiles.}
\label{fig:Stokes_plot}
\end{figure*}

The amplitude of the 
Q/I signal peak in the Sr~{\sc i} line at every spatial 
position is calculated using a Gaussian fit to each of the 140 profiles. 
Figure~\ref{fig:Stokes_plot} shows Q/I profiles obtained from a single pixel position 
(arbitrarily chosen at 9" spatial position) for $\mu$ = 0.2 to 0.8, corresponding to the Stokes Q/I images 
shown in Fig.~\ref{fig:stokes_image}. The red lines are the Gaussian fits to the profiles. We used the GAUSSFIT function available in 
the interactive data language to fit the Q/I profiles. 
An estimate of the 1-sigma error of the returned parameters for each fit was also calculated by the fitting
procedure.
The fitted amplitudes and error are stored for further investigations. 
Our obtained Q/I amplitudes are consistent with those obtained by \cite{Stenflo1997}. 
Clear spatial variations of the amplitudes of the Q/I peaks can be appreciated at different $\mu$ positions under adequate seeing conditions. 
Such variations are  of solar origin and are no longer detectable when seeing conditions deteriorate give that 
granular regions and intergranular lanes cannot be easily distinguished. 

\subsection{Correlations of Q/I with the continuum intensity}\label{sec:sec4_3}

\begin{figure*}[ht]
\centering
{
\includegraphics[width=0.5\textwidth,clip=]{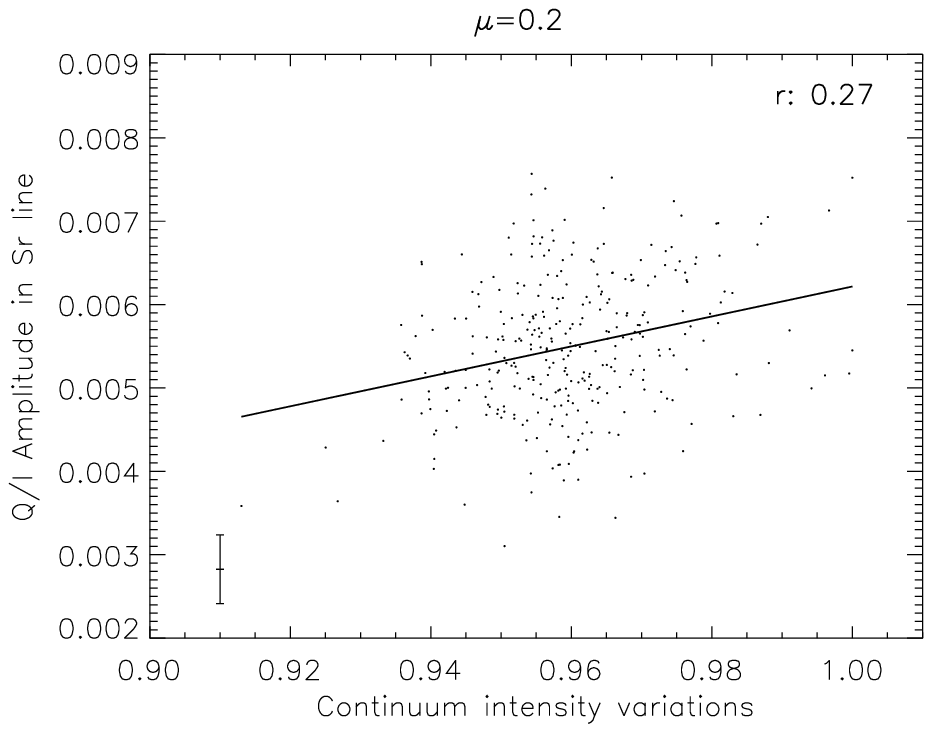}
   \hspace*{-0.02\textwidth}
 \includegraphics[width=0.5\textwidth,clip=]{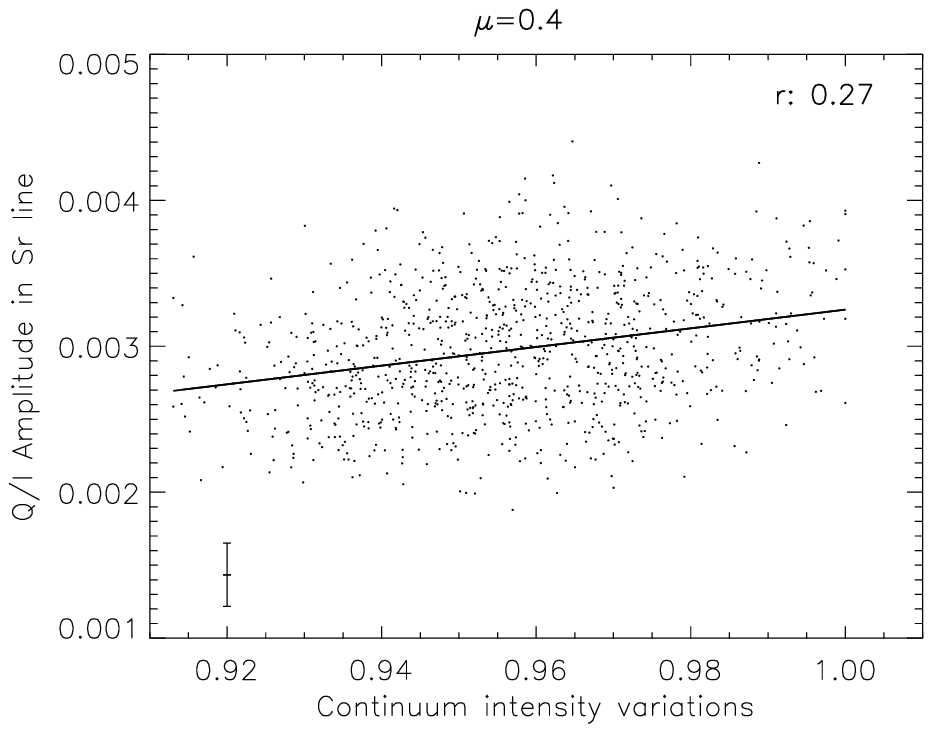}}
 {\includegraphics[width=0.5\textwidth,clip=]{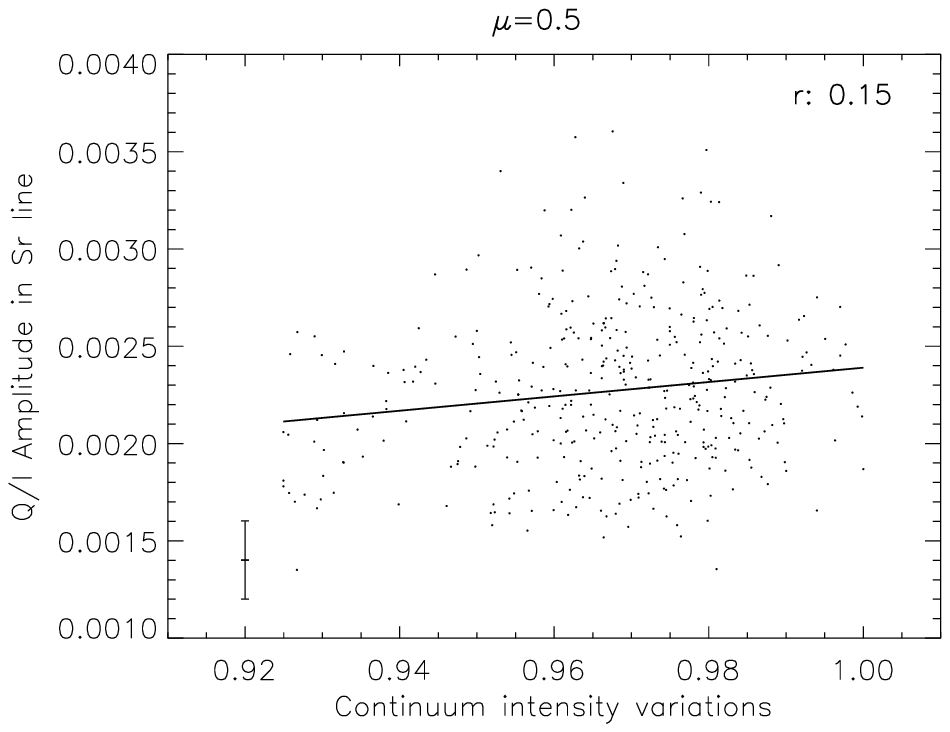}
\hspace*{-0.02\textwidth}
 \includegraphics[width=0.5\textwidth,clip=]{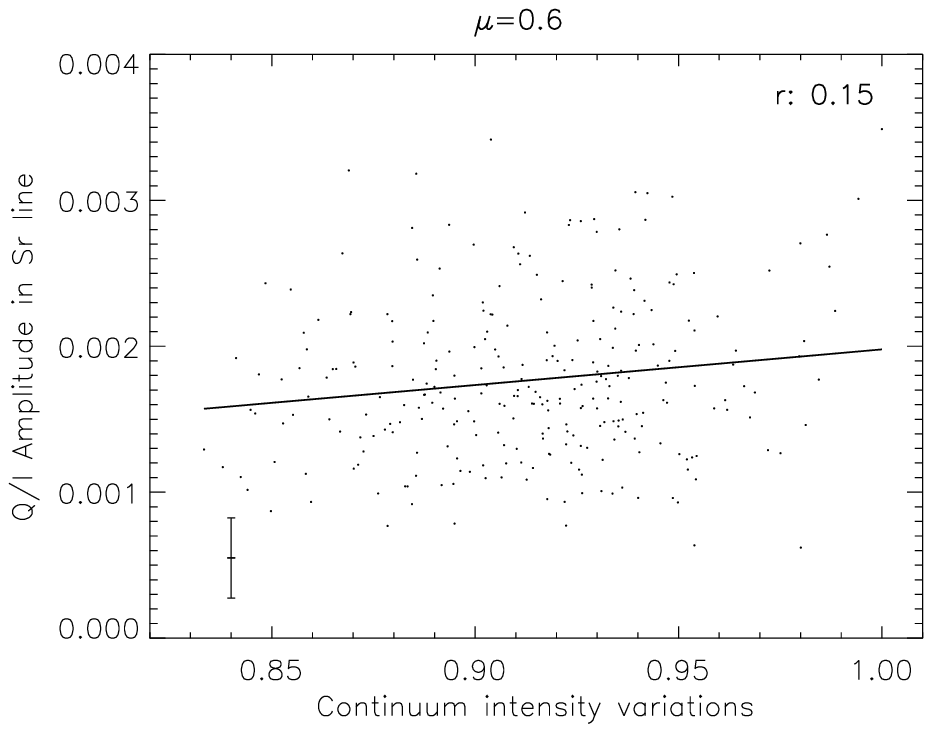}}
  {\includegraphics[width=0.5\textwidth,clip=]{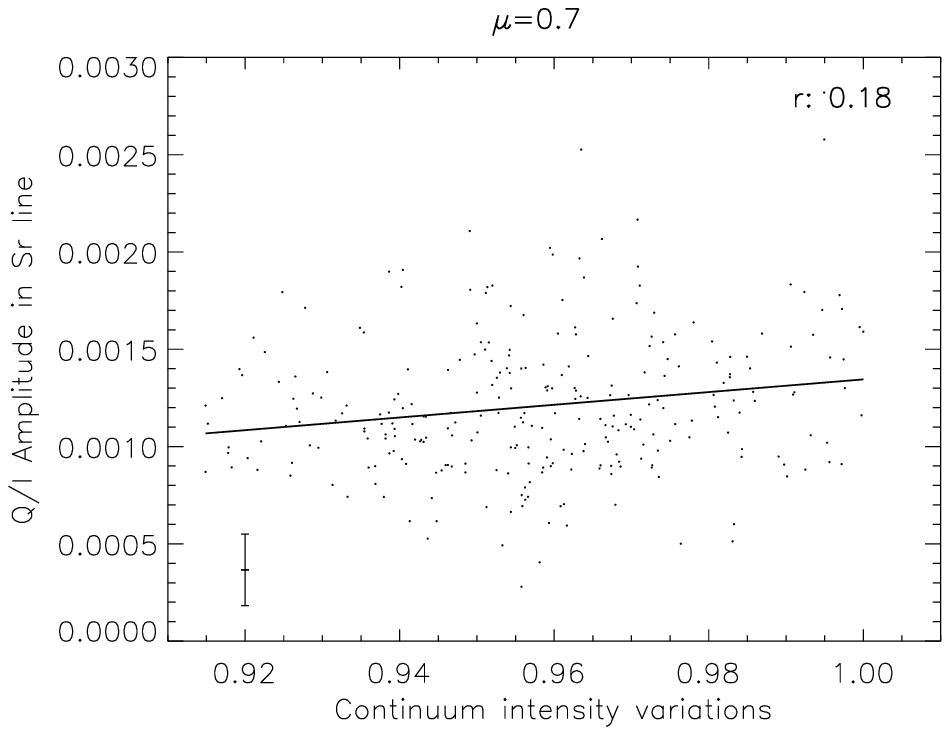}
\hspace*{-0.02\textwidth}
 \includegraphics[width=0.5\textwidth,clip=]{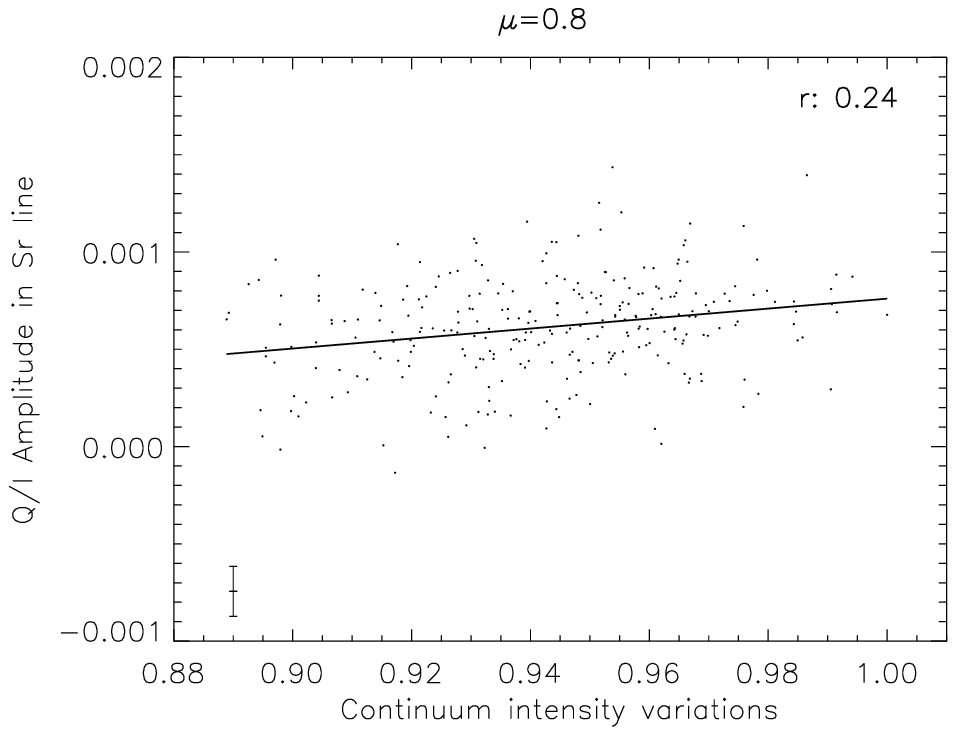}}
 \vspace{0.005\textwidth}
\caption{Scatter plot relating amplitudes of Sr~{\sc i}~ 4607~\AA~Q/I peak to normalized continuum intensity 
 at different $\mu$. The solid line represents a linear regression of the data. Its positive slope indicates larger 
 polarization in the granulation. The Pearson correlation coefficient ({\it r}) 
value  is annotated in each graph. The estimated error at all points  is given by the error bar in the bottom left corner of each plot.
}
\label{fig:scatter_plot}
\end{figure*}

\begin{figure*}
\centering
{
\includegraphics[width=0.78\textwidth,clip=]{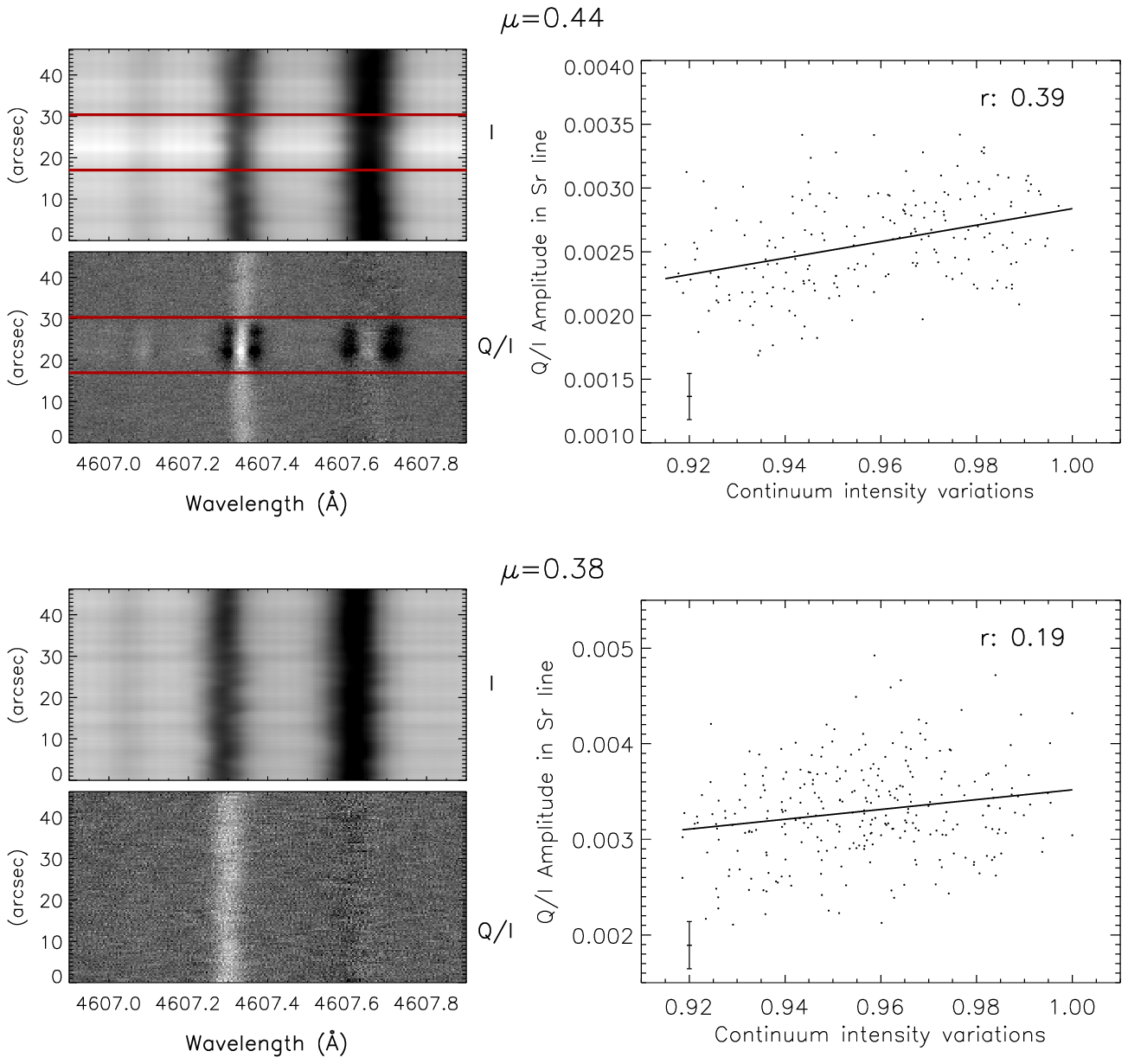}
 }
 \vspace{0.0\textwidth}
\caption{Left: Averaged Stokes images (I and Q/I) at $\mu$=0.44 (upper panel) and 0.38 (lower panel).
Right: Scatter plots relating amplitudes of Q/I peak signals to continuum intensity.
The Pearson correlation coefficients ({\it r}) are reported
in the scatter plots. In the measurements taken at $\mu$ = 0.44, the active region is excluded for the calculation of {\it r}.
 Two red horizontal lines show the excluded spatial portion in the top left Stokes I and Q images.
The estimated error for each point is shown in the bottom left of each plot.}
\label{fig:scatter_plot1}
\end{figure*}

The scatter plots shown in Fig.~\ref{fig:scatter_plot} allow us to study the correlations between the polarimetric Q/I peak 
signals of the Sr~{\sc i} 4607~\AA~line (y-axis) and the normalized continuum intensity (x-axis), with each panel 
corresponding to a different $\mu$ between 0.2 and 0.8. 
The continuum intensities are obtained from the space-time maps as described in section~\ref{sec:sec4_1}.
The error in the Q/I peak amplitudes is estimated  from the Gaussian fit procedure applied to the
profile. The solid line in the plot is a 
linear regression (obtained with the  linfit function available in the interactive data language). 
Figure~\ref{fig:scatter_plot1} shows
the Stokes I, Q/I images and scatter plots measured at the east limb 100$^"$ ($\mu$=0.44) and 74$^"$ ($\mu$=0.38) inside the 
solar limb. 
In the measurements taken at $\mu$ = 0.44, part of the slit is over an active region; such 
spatial positions were excluded  (two red horizontal lines showing the excluded spatial portion in the Stokes I and Q images)
from the calculation of the correlation coefficient.
For all the measurements the obtained slope of the linear regression is positive,
suggesting a trend of increasing linear polarization in the higher intensity regions.
The Pearson correlation  coefficients ({\it r}) obtained from each of the scatter plots, are shown both in the plots 
themselves and in Table ~\ref{Table:1}. 
 The obtained correlation coefficients are positive 
for all $\mu$.
 Furthermore, the statistical significance of the obtained coefficients ({\it r}) were evaluated by determining the {\it p}-values. 
The {\it p}-value is the probability of finding the corresponding {\it r}-coefficients provided the 
null hypothesis (i.e., the lack of correlation 
between the intensity and the Q/I peak amplitudes) is true. 
  As such it is worth noting that having obtained low {\it p}-value only serves to discard the null hypothesis, but cannot be 
taken as confirmation of a positive correlation.
The calculated {\it p}-values for all the measurements  are
given in the fourth column of Table~\ref{Table:1}.
For most of the measurements, 
the obtained {\it p}-values are found to be less than 10$^-$$^3$, which corresponds to a 
confidence level of  99.9\% for rejecting the null hypothesis.

\begin{table}[h!]
\centering
\footnotesize

\caption{Obtained Pearson correlation coefficient ({\it r}), probability {\it p}-value with total number of points ({\it N}) 
from  scatter plot at different $\mu$.}
\begin{tabular}{|l|l|l|l|}
 \hline
&&&\\
           
  Limb                     & {\it N}  &   {\it r}      &  {\it p}-value \\
  distances ($\mu$)   &          &                &    (approximate)\\
\hline  
&&&\\
 0.2        & 328 & 0.274  &  4.7$\times$10$^{-7}$ \\
 0.38       & 280 & 0.197  &  9.2$\times$10$^{-4}$  \\

 0.4      & 980 & 0.266  &  6.7$\times$10$^{-16}$  \\
 
 0.44      & 280 & 0.386  &1.5 $\times$10$^{-8}$ \\
 
 0.5        & 420 & 0.147  & 2.5$\times$10$^{-3}$   \\
 0.6        & 280 & 0.149  & 1.2$\times$10$^{-2}$  \\
 0.7        & 280 & 0.179  & 2.6$\times$10$^{-3}$  \\
 0.8  e     & 280 & 0.237  & 6.3$\times$10$^{-5}$   \\
\hline
\end{tabular}
\label{Table:1}
\end{table}

\section{Discussion and conclusion}\label{sec:sec5}

We performed scattering polarization measurements of the Sr {\sc i}
4607.3~\AA~line at different limb distances with ZIMPOL at the GREGOR telescope in Tenerife. 
The measurements were taken between $\mu$= 0.2 and 0.8 on the solar disk  for different time duration 
depending upon the seeing condition (see Table~\ref{Table:obs1}). The direction of 
positive Stokes Q was always parallel to the nearest solar limb.
 We averaged a few subsequent frames over time for all the measurement to improve the S/N in the Stokes images. The number 
of frames averaged (total integration time) at each $\mu$ are different (see Table~\ref{Table:obs2}), depending upon the signal obtained 
at that position.
We find spatial variations of
the Q/I scattering polarization signal of the Sr {\sc i} 4607~\AA~ line at all measured positions.
 These obtained spatial variations of
the Q/I scattering polarization signal are not directly comparable among all $\mu$.
These results were obtained with a spatial resolution between $\sim$0.66" and 0.99".
The spatial scale of the observed variations
is comparable to the granular scale. 
We find a positive correlation between the scattering polarization peak 
amplitudes of the Sr {\sc i} 4607~\AA~line and the continuum
intensity at every measured position (as shown in Table~\ref{Table:1}).  
This implies, statistically, that the polarization inside granular regions is higher than in 
the intergranular lanes. The positive correlation obtained in this work supports
the results reported independently by \cite{Malherbe2007} and
\cite{Bianda2018}, obtained in both cases at $\mu$ = 0.3 using a spectrograph.

The positive correlation found here seemingly contradicts
the negative correlation reported by \cite{Zeuner2018}, obtained at $\mu$ = 0.6 with a spatial 
resolution of 0.16  pixel$^-$$^1$ and spectral resolution 25~m\AA~with the FSP mounted on filter-graph
TESOS at the VTT in Tenerife.  They acquired two dimensional Stokes images 
(FOV 20"$\times$20") with 2.5 s integration time per wavelength position, and the noise in the linear Stokes 
 is 0.3$\%$. In our measurement, the acquired Stokes Q/I image has a noise level of 0.22$\%$ for a 
 total integration time of 2.9 min with a 
 spatial resolution of $\sim$ 0.66" pixel$^-$$^1$ and spectral resolution 10 m\AA.
The correlation coefficient obtained by \cite{Zeuner2018} is r = -0.17 at $\mu$ = 0.6, while our obtained result is r = +0.15
at the same position.
Their observation supports the 
simulation by \cite{Aleman2018}, that foresees an anti-correlation when assuming a high spatial resolution (0.1''). 
By numerically deteriorating the S/N and the spectral and spatial
resolutions of the simulated observations, \cite{Aleman2018} showed that it is possible to reproduce the positive correlation.
Their theoretical study also suggests that, for 
investigating the scattering polarization signals of the Sr {\sc i} 4607~\AA~line, a better instrument would be a
2D spectropolarimeter with a spectral resolution not worse than 20 m\AA, a spatial resolution $\sim$0.1'' and a polarimetric
sensitivity better than 10$^-$$^4$.
 In our measurement, the spectral resolution is better than the setup used by \cite{Zeuner2018}, although the spatial resolution is poor.
We attribute the difference between the 
sign of the correlation found by us and the one
reported by \cite{Zeuner2018}  to the lower spatial resolution, longer integration time, and different S/N of our observations.

The observed results presented here are mostly limited by 
seeing conditions and by the available photon statistics. 
A significant improvement of this study can be achieved by performing imaging polarimetric
measurements at the line core of the Sr {\sc i} line with high spatial resolution in a large telescope such 
as Daniel K. Inouye Solar Telescope.
In the future we intend to develop a Fabry-Perot filter-based polarimeter system for conducting synoptic
measurements of the Sr {\sc i} 4607~\AA~ line, with the goal of investigating the subgranular photospheric 
magnetic field. This system could be installed as a second generation instrument on Daniel K. Inouye Solar Telescope.

\begin{acknowledgements}
IRSOL is supported by the Swiss Confederation (SEFRI),
Canton Ticino, the city of Locarno and the local municipalities.
This research work was financed by SNF 200020\_169418. The 1.5-meter GREGOR solar telescope was built by a German 
consortium under the leadership of the Kiepenheuer-Institut fur Sonnenphysik 
in Freiburg with the Leibniz-Institut f\"ur Astrophysik Potsdam, the Institut f\"ur Astrophysik G\"ottingen, 
and the Max-Planck-Institut f\"ur Sonnensystem forschung in G\"ottingen as partners, and with contributions by the Instituto de
Astrofsica de Canarias and the Astronomical Institute of the Academy of Sciences of the Czech Republic.
We thank the referee for insightful
comments which helped us to improve the content in the
manuscript.
\end{acknowledgements}

   \bibliographystyle{aa}
  \bibliography{sr_aa}

\end{document}